\documentclass{scsPaperFormattingTemplate-LaTex-Revised20150804}
\copyrightnotice{
	SpringSim-HPC 2016 April 3-6, Pasadena, CA, USA
	
	\copyright 2016 Society for Modeling \& Simulation International (SCS) 
}

\usepackage{balance}		
\usepackage{graphics}		
\usepackage{times}			
\usepackage{url}			
\usepackage{dblfloatfix}	

\usepackage{cite}
\usepackage[caption=false,font=footnotesize,labelfont=sf,textfont=sf]{subfig}
\usepackage{algorithm}
\usepackage{algorithmic}
\makeatletter
\def\url@leostyle{%
  \@ifundefined{selectfont}{\def\UrlFont{\sf}}{\def\UrlFont{\small\bf\ttfamily}}}
\makeatother
\def\pprw{8.5in}
\def\pprh{11in}

\usepackage[pdftex]{hyperref}
\hypersetup{
pdftitle={SCS Conference Proceedings Format},
pdfauthor={LaTeX},
pdfkeywords={SCS, proceedings, archival format},
bookmarksnumbered,
pdfstartview={FitH},
colorlinks,
citecolor=black,
filecolor=black,
linkcolor=black,
urlcolor=black,
breaklinks=true,
}

\begin{document}

\title{Managing Deadline-constrained Bag-of-Tasks Jobs on Hybrid Clouds}


\numberofauthors{2}
\author{
  \alignauthor Bo Wang\\
    \affaddr{Department of Computer Science and Technology, Xi'an Jiaotong University}\\
    \affaddr{ Xi'an, China 710049 \&}\\
    \affaddr{SKL Computer Architecture, ICT, CAS}\\
    \email{wangbo2012@stu.xjtu.edu.cn}\\
    \alignauthor Ying Song\\
    \affaddr{SKL Computer Architecture, ICT, CAS}\\
    \affaddr{Beijing, China 100190}\\
    \email{songying@ict.ac.cn}\\
    \alignauthor Yuzhong sun\\
    \affaddr{SKL Computer Architecture, ICT, CAS}\\
    \affaddr{Beijing, China 100190}\\
    \email{yuzhongsun@ict.ac.cn}\\
    \alignauthor Jun Liu\\
    \affaddr{SPKLSTN Lab, Department of Computer Science and Technology, Xi'an Jiaotong University}\\
    \affaddr{ Xi'an, China 710049}\\
    \email{liukeen@mail.xjtu.edu.cn}
}

\maketitle

\begin{abstract}
Outsourcing jobs to a public cloud is a cost-effective way to address the problem of satisfying the peak resource demand when the local cloud has insufficient resources. In this paper, we study on managing deadline-constrained bag-of-tasks jobs on hybrid clouds. We present a binary nonlinear programming (BNP) problem to model the hybrid cloud management where the utilization of physical machines (PMs) in the local cloud/cluster is maximized when the local resources are enough to satisfy the deadline constraints of jobs, while when not, the rent cost from the public cloud is minimized. To solve this BNP problem in polynomial time, we proposed a heuristic algorithm. Its main idea is assigning the task closest to its deadline to current core until the core cannot finish any task within its deadline. When there is no available core, the algorithm adds an available PM with most capacity or rents a new VM with highest cost-performance ratio. Extensive experimental results show that our heuristic algorithm saves 16.2\%-76\% rent cost and improves 47.3\%-182.8\% resource utilizations satisfying deadline constraints, compared with first fit decreasing algorithm.
\end{abstract}

\keywords{Bag-of-tasks; cloud computing; hybrid cloud; resource management; task scheduling}

\section{Introduction}\label{sec:introduction}

Hybrid cloud, combining a local cloud (private cloud) and a public cloud, is a cost-efficient way to address the problem of insufficient resources in the local cloud when its users have a peak resource demand as the peak load is much larger than average, but transient \cite{LoadBursting}. Surveyed by the European Network and Information Security Agency (ENISA), most of the small to medium enterprises prefer a mixture of cloud computing models (public cloud, private cloud) \cite{SMEs}.

Reducing total capital expenditure on resources is a main objective on hybrid clouds for a provider owning local resources. Generally, using the resources of the local cloud is costless or cheaper, considering that investment costs for the physical infrastructures are ``sunk costs'', compared with leasing the resources from a public cloud. Thus minimizing the costs for a private cloud provider on a hybrid cloud is the integration of maximizing the resource utilizations of the local cloud and minimizing the rent cost from the public cloud.

There are various researches on scheduling the scientific computing applications on hybrid clouds. A few works focus on minimizing the rent cost with deadline constraints \cite{BOT1,BOT2,BOT3,HCOC1,HCOC2,UCC13,reusable} or minimizing the makespan \cite{FermiCloud,CBS1,CBS2} for scientific computing applications by deciding which tasks should be outsourced to the public cloud. While, these works do not consider how the local cloud/cluster provisions resources, i.e. they do not provide the mapping between physical machines (PM) and provisioned resources in the local cloud. 

Only a few hybrid cloud managements \cite{Aneka1,Aneka2,Aneka3} have coordinated dynamic provisioning and scheduling that is able to cost-effectively complete applications within their respective deadlines. While these works separately scheduled tasks and provisioned resources, i.e., they first decided how many resources, each of which is either a VM in public cloud or a PM in local cloud(s)/cluster(s), used for running tasks and then provisioned the resources from the resource pool, considering that all of the resources are homogeneous.

Different from these existing works, we study on cost-efficiently mapping the tasks to the resources for deadline-constrained Bag-of-Tasks (BoT) jobs, a kind of very common application in the parallel and distributed systems \cite{common1,common2}, such as parallel image rendering, data analysis, and software testing \cite{BoTimage, BoTdata, BoTtest}, on a hybrid cloud with heterogeneous local resources. BoT jobs are often composed of hundreds of thousands of independent tasks and are CPU-intensive. 

In this paper, we model the task and resource managements of hybrid clouds into a binary nonlinear programming (BNP) model. In this model, the utilization of local resources is maximized while the cost for the resources leased from the public cloud is minimized. As BNP is NP-hard problem \cite{NIP}, we propose a heuristic algorithm to solve this BNP problem. In brief, the contributions of this paper can be summarized as follows:

\begin{enumerate}
\item We model the hybrid cloud management which maximizes the utilization of local resources and minimizes the cost for renting the resources of the public cloud for BoT jobs with deadline constraints into a BNP problem.
\item To solve the BNP problem in polynomial time, we propose a heuristic algorithm. The algorithm assigns a task to a core such that the difference between the task's finish time and its deadline is minimum in all assignments between unassigned tasks and cores of used PMs. If there is no such assignment, which means that there is no task can be completed within the deadline by the resources already used, the algorithm adds an available PM with most capacity or leases a VM with most cost-performance ratio from the public cloud when there is no available PM in the local cloud, and assigns tasks to the cores of the added PM/VM as the previous step.
\item We conduct extensive simulation experiments using two real work traces to investigate the effectiveness and efficiency of the proposed algorithm. The experiments results show that our heuristic algorithm saves 16.2\%-76\% cost for finishing jobs within their respective deadlines and improves 47.3\%-182.8\% resource utilizations, compared with first fit decreasing (FFD) algorithm.
\end{enumerate}

The rest of the paper is organized as follows. Section \ref{management} presents our model and the heuristic algorithm for solving the model. Section \ref{ERA} evaluates our work. Section \ref{RW} discusses related work and Section \ref{CFW} concludes this paper.

\section{Hybrid Cloud Management}\label{management}

\begin{figure}[!t]
\centering
\includegraphics[width=2.4in]{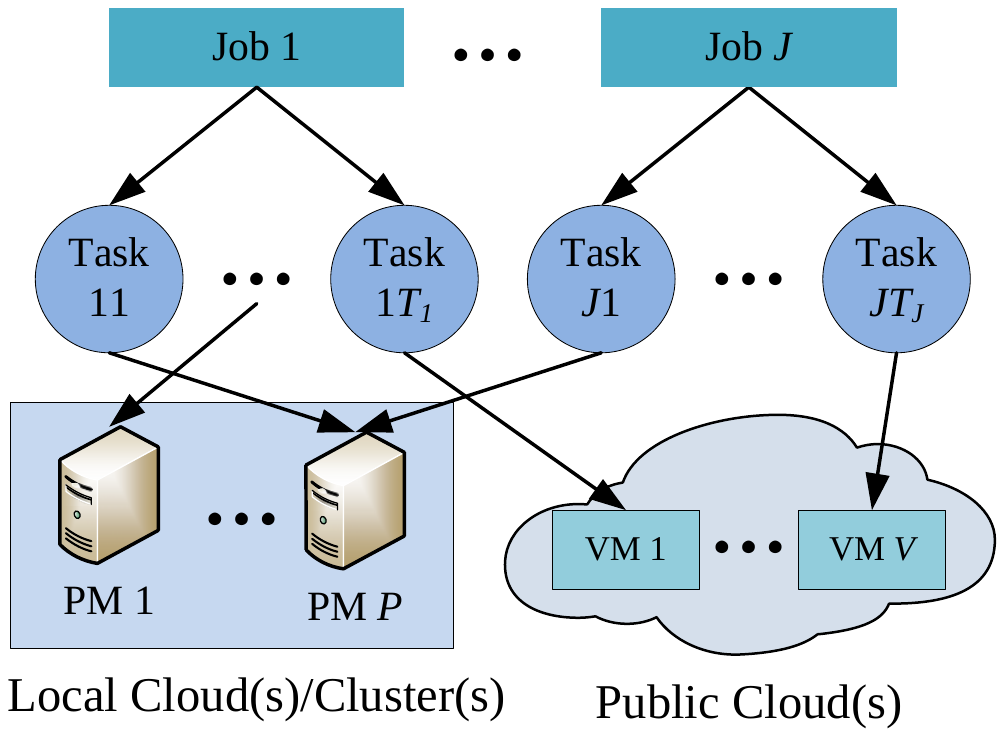}
\caption{Hybrid cloud environment. A task of jobs is assigned to either a PM in local cloud(s)/cluster(s) or a VM leased from public cloud(s).}
\label{arch}
\end{figure}

In a hybrid cloud, as shown in Fig.~\ref{arch}, a task of jobs runs on a core of a PM on the local cloud/cluster or of a VM leased from the public cloud. In this paper, the objective is to cost-efficiently provision resources to tasks and assign the tasks to the resources to meet the complete time within corresponding deadline in the hybrid cloud environment, i.e., to provide the mapping between the PMs or rented VMs and the tasks with minimal cost while fitting deadlines.

\subsection{Problem Formulation}
\label{PF}

We consider a hybrid cloud consisted of a local cloud/cluster and a public cloud. Multiple public clouds in a hybrid cloud can be seen as one big public cloud including the resources provisioned by these public clouds. Table~\ref{notations} summarizes notations used in this paper.

\begin{table}[!t]
\small
\renewcommand{\arraystretch}{1.2}
\centering
\begin{tabular}{|c|p{2.4in}|}
\hline
\textbf{Notations} & \textbf{Description} \\\hline
$J$ & The number of jobs running on the hybrid cloud.\\\hline
$T_i$ & The number of tasks constituting job $i$.\\\hline
$T$ & The number of tasks, $T = \sum_{i=1}^JT_i$.\\\hline
$t_{i,j}$ & The $j$th task of job $i$.\\\hline
$d_i$ & The deadline of job $i$.\\\hline
$r_{i,j}$ & The resource amount needed to complete $t_{i,j}$.\\\hline
$P$ & The number of PMs in the local cloud.\\\hline
$V$ & The maximal number of VMs leased from the public cloud.\\\hline
$p_k$& The price per unit time of container $k$ representing a VM provisioned by the public cloud when $P+1\leq k\leq P+V$.\\\hline
$c_k$& The cost for renting container $k$ ($P+1\leq k\leq P+V$).\\\hline
$N_k$ & The number of cores on container $k$.\\\hline
$r_k$ & The capacity of each core on container $k$.\\\hline
$\varepsilon$& A small positive constant such that $\varepsilon < \min\limits_{P+1\leq k\leq P+V} p_k/P$.\\\hline
$x_{i,j,k,l}$ & The binary variable representing whether $t_{i,j}$ is assigned to core $l$ of container $k$.\\\hline
$N_{RV}$ & The number of rented VMs.\\\hline
\end{tabular}
\caption{Notations.}
\label{notations}
\end{table}

There are $J$ jobs running on the hybrid cloud. Job $i$ ($i = 1,...,J$) is composed of $T_i$ independent tasks, $\{t_{i,j}\ |\ j=1,...,T_i\}$. $T=\sum_{i=1}^JT_i$ represents the number of all tasks. It needs $r_{i,j}$ resource amounts to complete $t_{i,j}$. Job $i$ must be completed before $d_i$. Without loss of generality, we assume that $d_1 \leq d_2 \leq \cdots \leq d_J$.

In the local cloud/cluster, there are $P$ PMs. In the public cloud, at most $V$ VMs are rented. We consider a PM or a rented VM as a resource container. Without loss of generality, we assume that the first $P$ containers are PMs and the rest of $V$ containers are VMs. The price per unit time of container $k$ ($k=P+1,...,P+V$) is $p_k$. Container $k$ ($k=1,...,P+V$) has $N_k$ cores each of which has $r_k$ capacity. It takes $r_{i,j}/r_k$ time for completing $t_{i,j}$ when running on a core of container $k$. If all the tasks scheduled to a core can be completed within respective deadlines, respectively, they can run in sequential order by their deadlines in ascending fashion to meet the deadlines. Then the deadline constraints can be formulated as follow (noticing that $d_1 \leq d_2 \leq \cdots \leq d_J$):
\begin{eqnarray}
\nonumber && \sum_{i'=1}^{i}\sum_{j=1}^{T_{i'}}(x_{i',j,k,l}\cdot\frac{r_{i',j}}{r_k}) \leq d_{i},\\
&&\hspace{0.1in} \forall i=1,...,J,\  \forall l=1,...,N_k,\  \forall k=1,...,P+V,
\label{deadline}
\end{eqnarray}
where the binary variable $x_{i,j,k,l}$ ($i=1,...,J$, $j=1,...,T_i$, $k=1,...,P+V$, $l=1,...,N_k$) represents whether $t_{i,j}$ is assigned to core $l$ of container $k$. If so, $x_{i,j,k,l}=1$, otherwise, $x_{i,j,k,l}=0$. The left sides of Inequalities~(\ref{deadline}) represent the finish times of $t_{i,j}$, $i=1,...,J$, $j=1,...,T_i$, respectively. The total time using a container is the maximum finish time of the task running on it, $\max\limits_{1\leq l\leq N_k}\sum_{i=1}^{J}\sum_{j=1}^{T_i}(x_{i,j,k,l}\cdot r_{i,j}/r_k)$, $k=1,...,P+V$. Thus, the costs for leasing VMs from public cloud respectively are
\begin{eqnarray}
\nonumber && c_k = \lceil\max\limits_{1\leq l\leq N_k}\sum_{i=1}^{J}\sum_{j=1}^{T_i}(x_{i,j,k,l}\cdot r_{i,j}/r_k)\rceil\cdot p_k,\\
&&\hspace{1.3in} \forall k=P+1,...,P+V,
\label{cost}
\end{eqnarray}
where $\lceil X \rceil$ is the ceiling integer of $X$, and the utilizations of PMs respectively are
\begin{eqnarray}\label{uti}
\nonumber&&u_k = \left\{ \begin{array}{l}
0, \hspace{0.4in}\textrm{if $\sum_{l=1}^{N_k}\sum_{i=1}^{J}\sum_{j=1}^{T_i}x_{i,j,k,l}=0$}\\
\frac{\sum_{l=1}^{N_k}\sum_{i=1}^{J}\sum_{j=1}^{T_i}(x_{i,j,k,l}\cdot r_{i,j})}{N_k\cdot\max\limits_{1\leq l\leq N_k}\sum_{i=1}^{J}\sum_{j=1}^{T_i}(x_{i,j,k,l}\cdot r_{i,j})},\hspace{0.2in} \textrm{else}
\end{array} \right.,\\
&&\hspace{2in}\forall k=1,...,P.
\end{eqnarray}
The overall utilization of local resources (PMs) is $\sum_{k=1}^Pu_k/P$.

We formulate the problem of hybrid cloud management as a BNP as follows:

\begin{equation}
\textrm{Minimize}\ \ -\varepsilon\cdot\sum_{k=1}^Pu_k + \sum_{k=P+1}^{P+V}c_k, \label{objective}
\end{equation}

subject to:
\begin{eqnarray}
&& \sum_{k=1}^{P+V}\sum_{l=1}^{N_k}x_{i,j,k,l} = 1,\  \forall j=1,...,T_i,\ \forall i=1,...,J,\label{st0}\\
\nonumber &&\textrm{Inequalities~(\ref{deadline})},\\
\nonumber && \textrm{Equations~(\ref{cost})},\\
\nonumber && \textrm{Equations~(\ref{uti})}, \label{st2}\\
\nonumber && x_{i,j,k,l}\in \{0, 1\},\  \forall j=1,...,T_i,\ \forall i=1,...,J,\\
&&\hspace{0.8in} \forall l=1,...,N_k,\ \forall k=1,...,P+V,\label{st3}
\end{eqnarray}
The decision variables are $x_{i,j,k,l}$ ($j=1,...,T_i$, $i=1,...,J$, $l=1,...,N_k$, $k=1,...,P+V$). In the objective function (\ref{objective}), $\varepsilon$ is a small constant such that $0<\varepsilon<p_k/P, \forall k=P+1,...,P+V$, which guarantees that the value of the objective function is negative if there is no VM leased from public cloud while positive if there is at least one rented VM. Therefore, the objective of this model is maximizing the local resource utilization when the resources of the local cloud are enough to complete jobs within respective deadlines while minimizing the cost ($\sum_{k=P+1}^{P+V}c_k$) of rented VMs when the local resources are insufficient. Constraints (\ref{st0}) ensure that each task must be assigned to only one core.
 Constraints (\ref{st3}) represent the binary requirements for the decision variables. After solving this model, we achieve the task assignments, $x_{i,j,k,l}$ ($i=1,...,J$, $j=1,...,T_i$, $k=1,...,P+V$, $l=1,...,N_k$), and the renting time for each VM, $c_k/p_k$ ($k=P+1,...,P+V$).

\subsection{The Heuristic Algorithm}

As BNP is NP-hard \cite{NIP}, we propose a heuristic algorithm to solve the model presented in Section~\ref{PF} in polynomial time. The main idea of the algorithm is to assign the task to a core so that the finish time of the task is closest to its deadline. If there is no enough resource, the algorithm adds an available PM with most capacity or leases a VM with best cost-performance ratio from the public cloud when there is no available PM in the local cloud. The details of the algorithm are described as follows, outlined in Algorithm 1.

\begin{algorithm}[!t]\small
\caption{assigning tasks to PMs or/and rented VMs}
\renewcommand{\algorithmicrequire}{\textbf{Input:}}
\renewcommand\algorithmicensure {\textbf{Output:} }
\renewcommand{\algorithmiccomment}[1]{{\ /*\textit{#1}*/}}
\begin{tabular}{cp{0.395\textwidth}}
$\mathcal{T}$:&the set of unassigned tasks, 3-tuples: (an unassigned task, its requirement of resource amount, its deadline);\\
$\mathcal{C}$:&the set of available cores, 4-tuples: (a core, its capacity, the finish time of all tasks assigned to it, the PM/VM containing it);\\
$\mathcal{PM}$:&the set of available PMs, 3-tuples: (an available PM, its core number, the capacity of a core);\\
$\mathcal{VT}$:&the set of available VM instance types, 4-tuples: (an available VM instance type, its core number, the capacity of a core, its price per unit time);\\
$\mathcal{A}$:&the set of assignments, 5-tuples: (a task, the core running it, the PM/VM containing the core, its start time).
\end{tabular}
 \\
\begin{algorithmic}[1]\small
\REQUIRE $\mathcal{T}$; $\mathcal{PM}$; $\mathcal{VT}$
\ENSURE $\mathcal{A}$
\WHILE{$\mathcal{T} \neq \phi$}
	\IF [There is no available resource] {$\mathcal{C} = \phi$}
		\IF [Adding the PM with maximum capacity] {$\mathcal{PM} \neq \phi$}
			\STATE $pm \gets p: (p\in\mathcal{PM})\wedge(p(2)\cdot p(3)=\max\limits_{p'\in\mathcal{PM}}(p'(2)\cdot p'(3)))$;\\\COMMENT{$p(i)$ is $i$th element in tuple $p$.}
			\STATE $\mathcal{PM} \gets \mathcal{PM}\setminus\{pm\}$;
			\FOR{$i=1$ to $pm(2)$}
				\STATE $\mathcal{C} \gets \mathcal{C}\cup\{($new core$, pm(3), 0, pm(1))\}$;
			\ENDFOR
		\ELSE [Renting an VM with highest cost-performance ratio]
			\STATE $vm \gets v: (v\in\mathcal{VT})\wedge$ C1$(v) \wedge$ C2$(v) \wedge$ C3$(v)$;\\/* C1$(v): (\forall t)((t\in \mathcal{T})\wedge(t(2)/v(3)\leq t(3)))$,\\C2$(v): v(2)\cdot v(3)/v(4)=$\\\hspace{0.8in}$\max\limits_{(v'\in\mathcal{VT})\wedge \textrm{\scriptsize C1}(v')}(v'(2)\cdot v'(3)/v'(4)))$,\\C3$(v): v(4)=\min\limits_{(v'\in\mathcal{VT})\wedge \textrm{\scriptsize C1}(v')\wedge \textrm{\scriptsize C2}(v')}{v'(4)}$ */
			\FOR{$i=1$ to $vm(2)$}
				\STATE $\mathcal{C} \gets \mathcal{C}\cup\{($new core$, vm(3), 0,$ a new VM instance of type $vm(1))\}$;
			\ENDFOR
		\ENDIF
	\ELSE [Scheduling a task on an available core]
		\STATE $(task, core) \gets (t, c): \textrm{C4}(t, c)\wedge\textrm{C5}(t, c)\wedge\textrm{C6}(t, c)$;\\ /* $\textrm{C4}(t, c): (t \in \mathcal{T}) \wedge (c \in \mathcal{C}) $\\\hspace{1.43in}$\wedge (t(3)\geq c(3)+t(2)/c(2))$,\\$\textrm{C5}(t, c): t(3)-(c(3)+t(2)/c(2))$\\\hspace{0.65in}$ = \min\limits_{\textrm{\scriptsize C4}(t', c')}{(t'(3)-(c'(3)+t'(2)/c'(2)))}$,\\$\textrm{C6}(t, c): t(2) = \max\limits_{\textrm{\scriptsize C4}(t', c')\wedge\textrm{\scriptsize C5}(t', c')}{(t'(2))}$ */
	    \IF{$\{(task, core)\} \neq \phi$}
	    	\STATE $\mathcal{A} \gets \mathcal{A}\cup\{(task(1), core(1), core(4), core(3)$\\\hspace{1.1in}$, core(3)+task(2)/core(2))\}$;
	    	\STATE $core(3) \gets core(3)+task(2)/core(2)$;
	    	\STATE $\mathcal{T} \gets \mathcal{T}\setminus \{task\}$;
	    \ELSE[There is no available resource in $\mathcal{C}$ for any task]
	    	\STATE $\mathcal{C} \gets \phi$;
	    \ENDIF
	\ENDIF
\ENDWHILE
\end{algorithmic}
\end{algorithm}

If there is no available resource (line 2), the algorithm would add an available PM (lines 3-8) or lease a VM (lines 9-14) from public cloud when there is no available PM in local cloud. The selection principle of an PM is to selecting the PM with most capacity (line 4). The selection principle of rented VM is that the selected VM has the capacity to complete any task when running alone (C1 in line 10) and has best cost-performance ratio (C2 in line 10). If there are multiple types of VMs having same cost-performance ratio, the algorithm selects the VM with minimal price per unit time (C3 in line 10). After selection, the algorithm adds the cores of selected PM (lines 6-8) or rented VM (lines 11-13) to the pool of available cores.

When there are one or more available cores (line 15), the algorithm assigns the unassigned tasks to these cores (lines 15-24). For all of assignments between unassigned tasks and available cores, the algorithm examines the finish times of the tasks and selects the assignment that the difference between the finish time of the task and its deadline is minimal and that the task is finished within its deadline (lines 16-20). If no assignment that the task can be finished within its deadline, there is no available resource for the unassigned task (lines 21-22).

The algorithm repeats the above steps until there is no unassigned task (line 1).

The computing resource consumed by the algorithm are mainly composed of the selection of an available PM or VM and the decision of an assignment between a task and a core. In real world, the numbers of VM instance types and cores in a PM/VM both are a few tens or fewer, thus the selection of a PM or VM and the decision of an assignment are $O(P)$ and $O(T)$, respectively, in time complexity. Therefore, assigning tasks to PMs is $O(T\cdot(P+T))$ in time complexity. We assume that there are $N_{RV}$ VMs leased from public cloud for completing the tasks within their respective deadlines. Assigning tasks to a VM is $O(T)$ in time complexity. Thus, Assigning tasks to the rented VMs is $O(T\cdot N_{RV})$ in time complexity. Hence, the algorithm is $O(T\cdot(P+T+N_{RV}))$ in time complexity, overall.

\section{Performance Evaluation} \label{ERA}

In this section, we introduce our testbed and experiment design, and then discuss the experimental results.

\subsection{Testbed and Experiments Design} \label{TED}

\begin{table}[!t]
\centering
\includegraphics{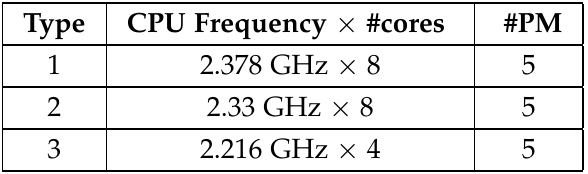}
\caption{The CPU configurations and numbers of PMs used for local resources.}
\label{PM}
\end{table}


We use a 3-month trace collected from the University of Luxemburg \textit{Gaia} cluster system and a 2-month trance collected from NASA Ames iPSC/860, i.e., \textit{UniLu Gaia log} and \textit{NASA iPSC log} in Parallel Workloads Archive \cite{PWA}, to evaluate the performance of our algorithm. We assume that the trace data are the information of tasks running on 1 GHz cores. We set the deadline of each job as $ \alpha $ (1, 2, 3, or 4) times of its run time on a 2 GHz core.

The PMs used as the local resources are shown in Table~\ref{PM}. In public cloud, we use a compute optimized instance type, c3.large in EC2 \cite{EC2}, because it has the best cost-performance ratio for CPU-intensive applications, compared with other instances provisioned by EC2. Each VM instance has 2 vCPUs with 2.7GHz. The price of a VM instance is \$0.105 per hour (in US east (N. Virginia)).

We use three performance metrics to evaluate task management algorithms:
\begin{itemize}
\item \textbf{\textit{rent cost}}: the cost for leasing VMs from the public cloud;
\item \textbf{\textit{makespan}}: the latest finish time of tasks;
\item \textbf{\textit{resource utilization}}: the overall utilization of used PMs and rented VMs.
\end{itemize}

\subsection{Comparison of Task Management Algorithms} \label{Comparison}

In this section, we compare our Heuristic Algorithm (HA) with First Fit Decreasing (FFD) \cite{FFD} in both performance and overhead. FFD is assigning the longest task to the first core on which it will fit. The time complexity of FFD is $O(T\cdot(P+N_{RV}))$ \cite{FFD} if the tasks have been sorted in their sizes. Before running FFD, we sort PMs on their capacities in descending order in advanced.

\subsubsection{Performance Comparison}\label{PC}

\begin{figure}[!t]
    \centering
    \subfloat[Gaia]{
    \includegraphics[width=0.45\columnwidth]{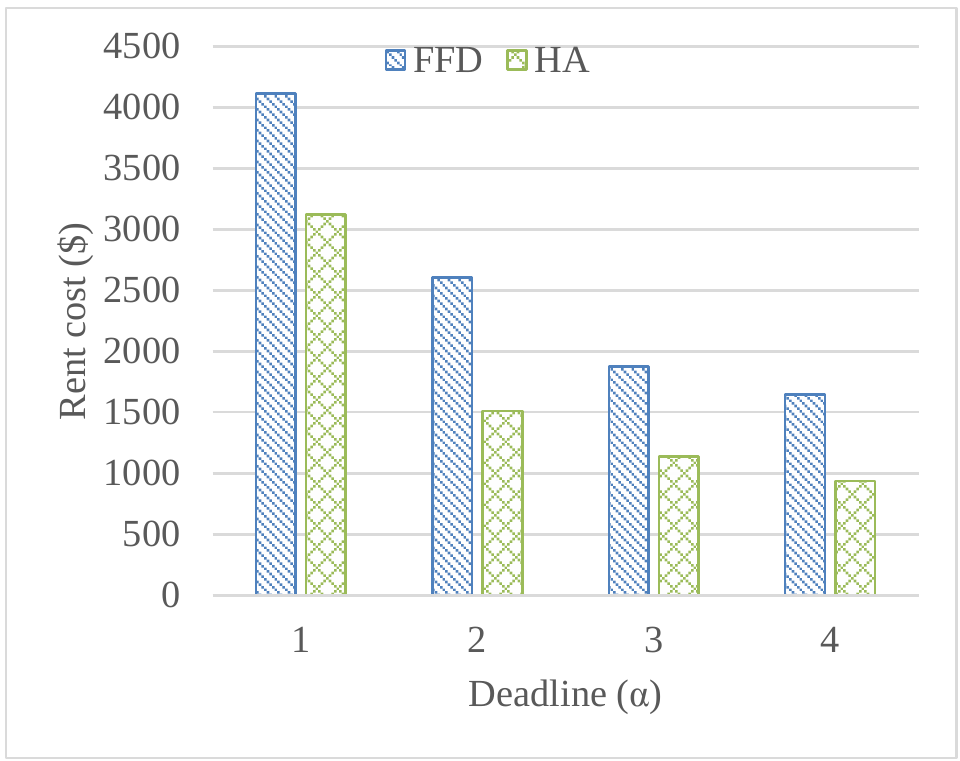}
    }
    \subfloat[NASA]{
    \includegraphics[width=0.45\columnwidth]{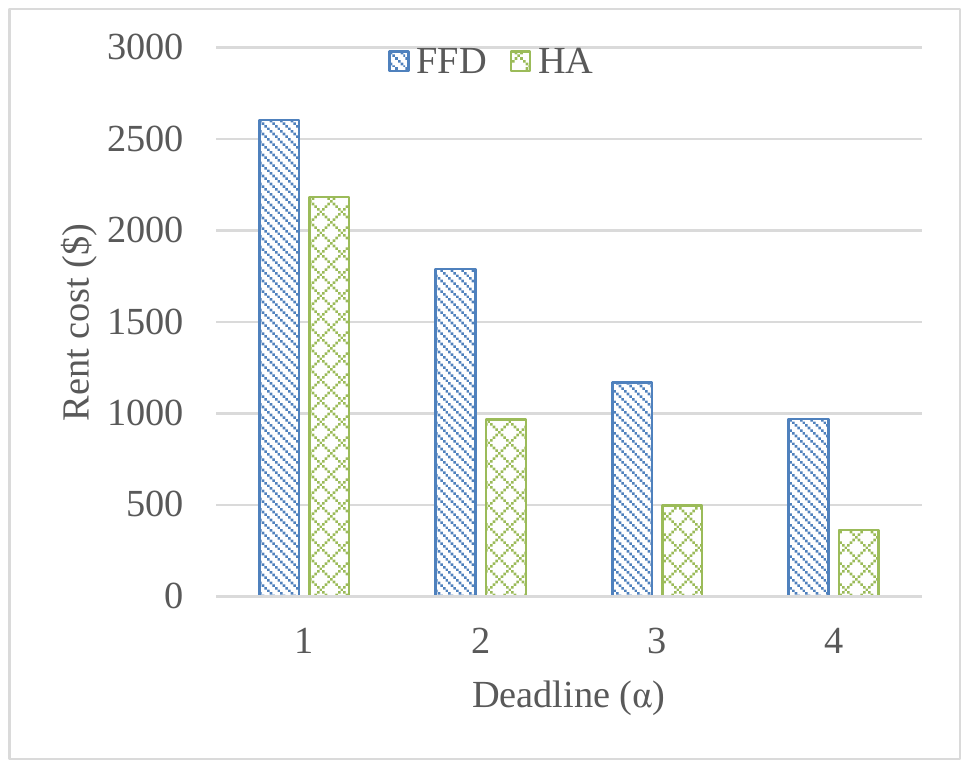}
    }
    \caption{{The rent costs for finishing Gaia (a) and NASA (b) tasks within their respective deadlines.}}\label{rentcost}
\end{figure}

\begin{figure}[!t]
    \centering
    \subfloat[Gaia]{
    \includegraphics[width=0.45\columnwidth]{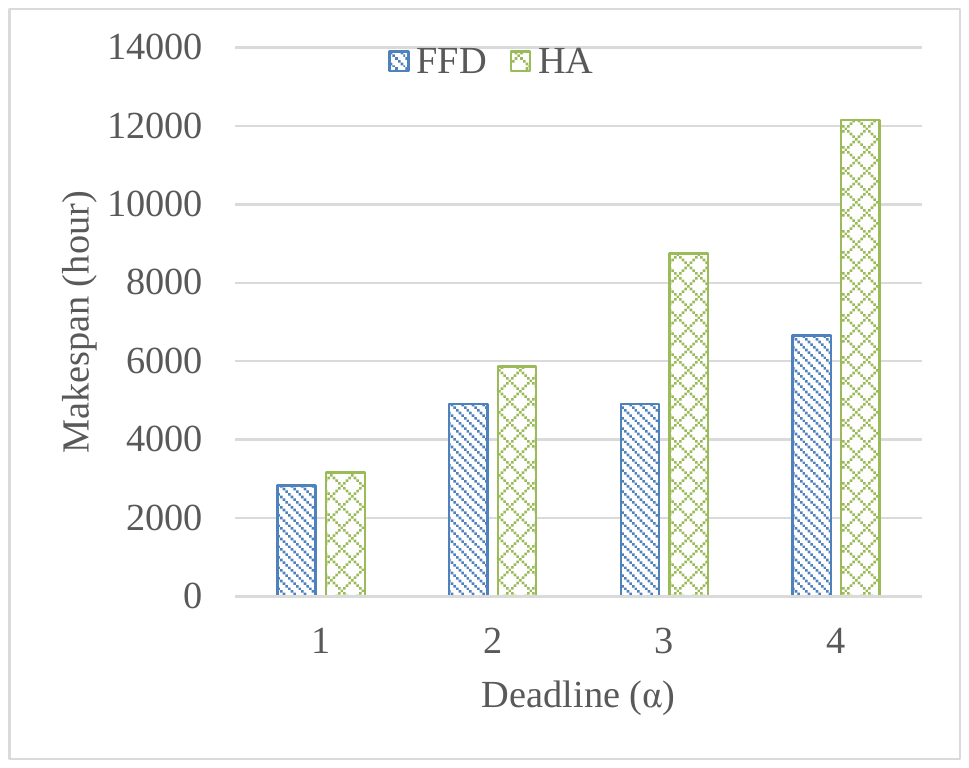}
    }
    \subfloat[NASA]{
    \includegraphics[width=0.45\columnwidth]{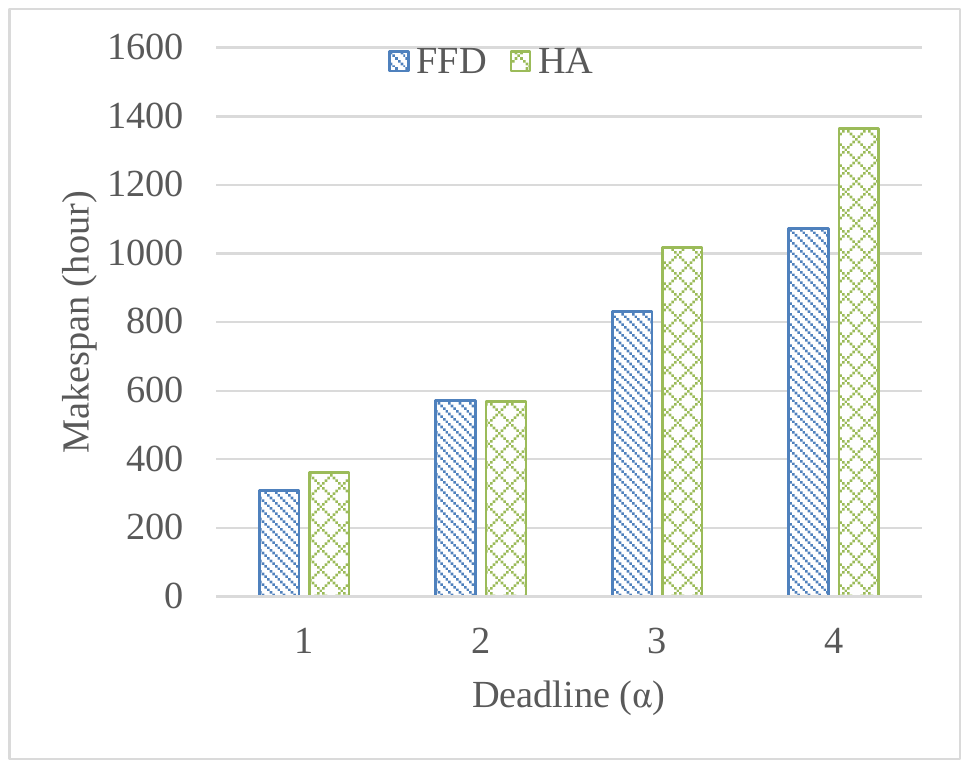}
    }
    \caption{{The time finishing Gaia (a) and NASA (b) tasks.}}\label{makespan}
\end{figure}

\begin{figure}[!t]
    \centering
    \subfloat[Gaia]{
    \includegraphics[width=0.45\columnwidth]{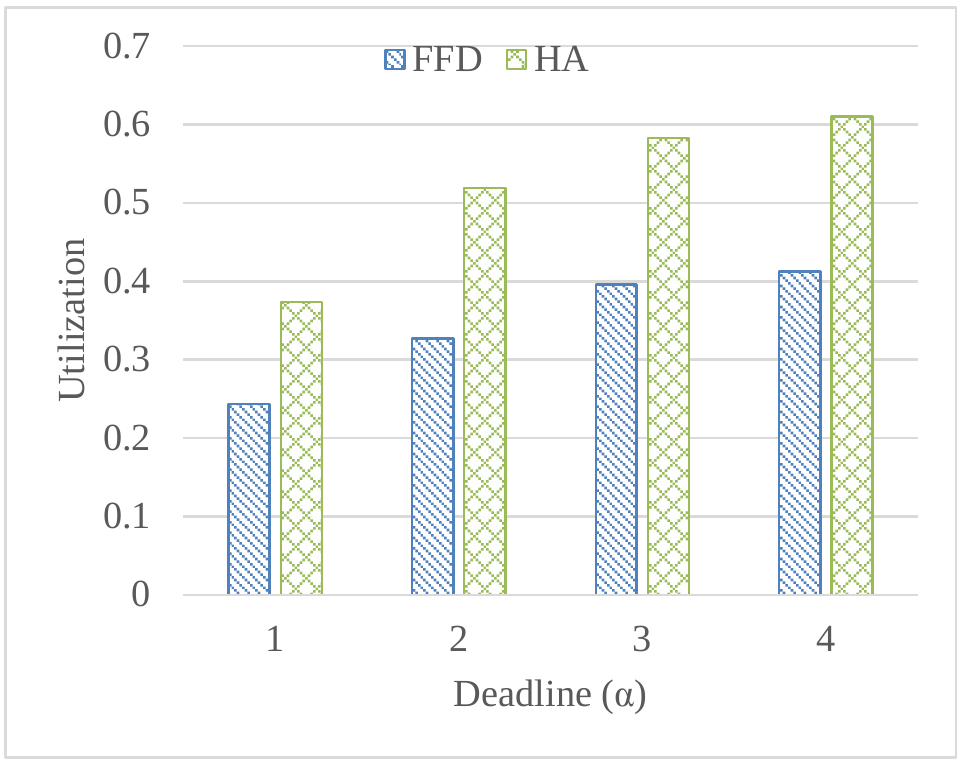}
    }
    \subfloat[NASA]{
    \includegraphics[width=0.45\columnwidth]{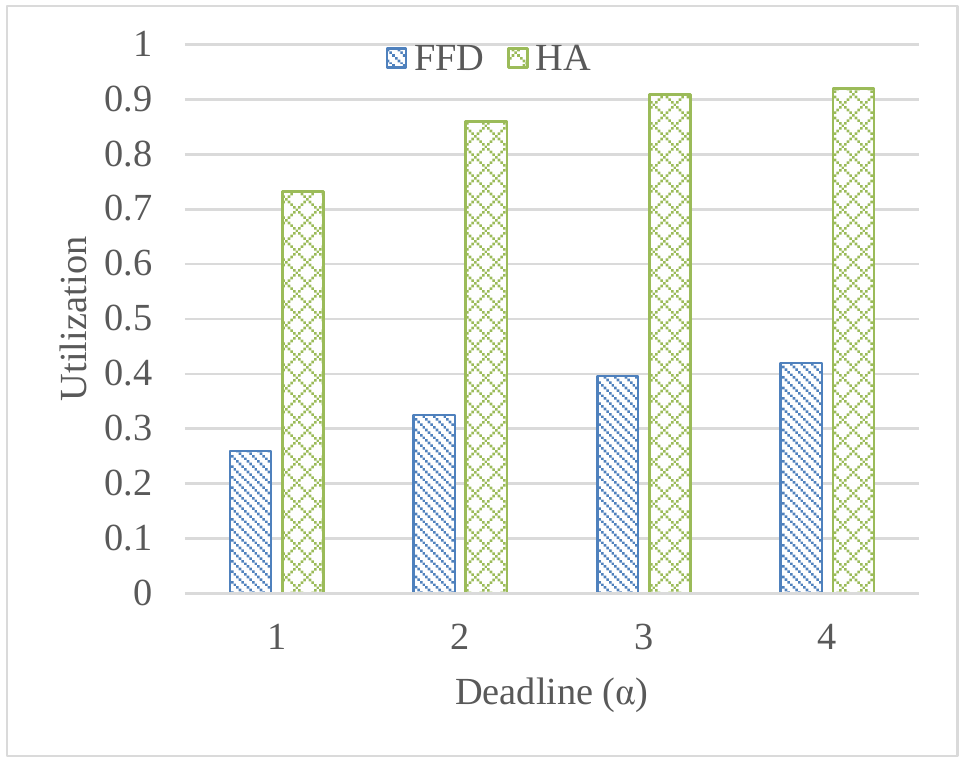}
    }
    \caption{{The utilization of PMs and VMs running Gaia (a) and NASA (b) tasks.}}\label{utilization}
\end{figure}

Figures~\ref{rentcost}, \ref{makespan}, and \ref{utilization} show the rent costs, the makespans of tasks, and the overall resource utilizations for finishing the tasks, managed by FFD and HA, within their respective deadlines, respectively.

For completing Gaia and NASA tasks within their respective deadlines, respectively, as shown in Fig.~\ref{rentcost}, HA consumes less 16.2\%-76\% cost than FFD for leasing VMs from public cloud. While HA postpones the finish time of these tasks to 11.8\%-82.6\% later, as shown in Fig.~\ref{makespan}, compared with FFD, with completing the tasks within their respective deadlines. The less rent cost and longer finish time of HA imply that HA makes better use of resources, compared with FFD. From Fig.~\ref{utilization}, we can see that the resource utilization of HA is 47.3\%-182.8\% more than that of FFD. These can be explained as follows. FFD schedules longest task first, and thus there will be small tasks left to be scheduled after assigning most of tasks. For finishing the left small tasks within their respective deadlines, only a few small tasks are assigned to one rented VM. Therefore, these VMs running only small tasks are under-utilised as rented VMs are charged by unit time, leading to high costs and low utilizations. While HA schedules the task closest to its deadline first, and thus rarely has the problem.

Thus, HA is much better than FFD in minimizing costs and improving resource utilizations in resource and task managements of hybrid clouds.

Figures~\ref{rentcost} and \ref{makespan} show that the rent cost is decreased with increasing the deadline while the finish time increases with deadline. The reason is that tasks are allocated less resources, which leads to longer waiting time of tasks and thus longer time to finish these tasks, if their deadlines are postponed.


\subsubsection{Overhead \& Scalability} \label{overhead}

In this section, we experimentally compare HA and FFD in overhead and scalability.

\begin{figure}[!t]
    \centering
    \subfloat[Gaia]{
    \includegraphics[width=0.45\columnwidth]{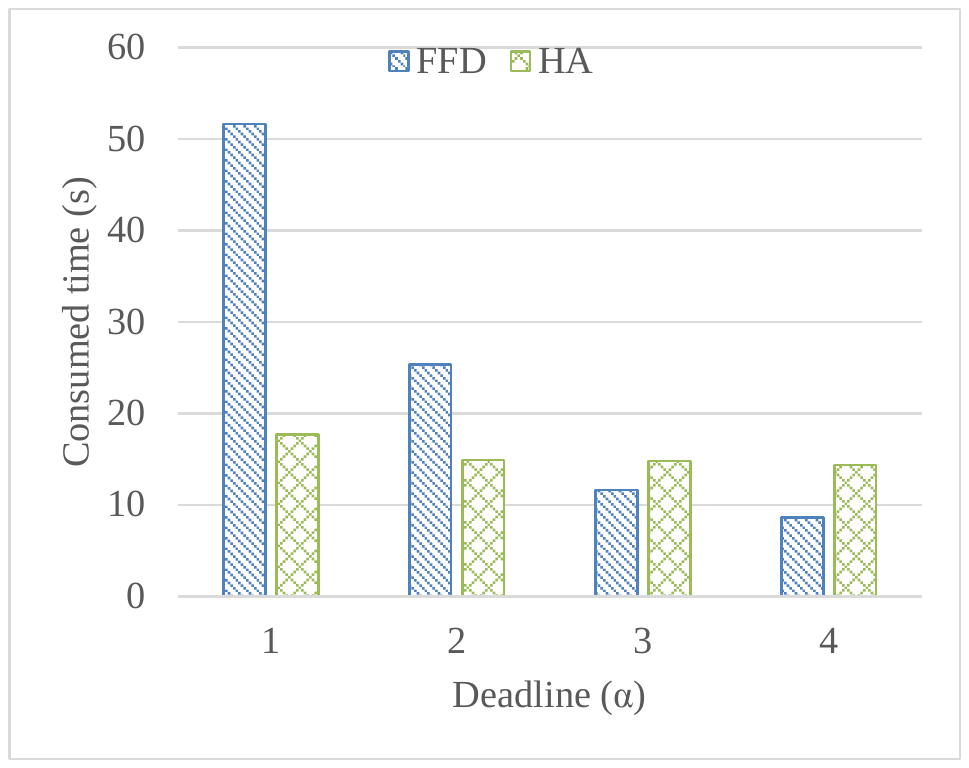}
    }
    \subfloat[NASA]{
    \includegraphics[width=0.45\columnwidth]{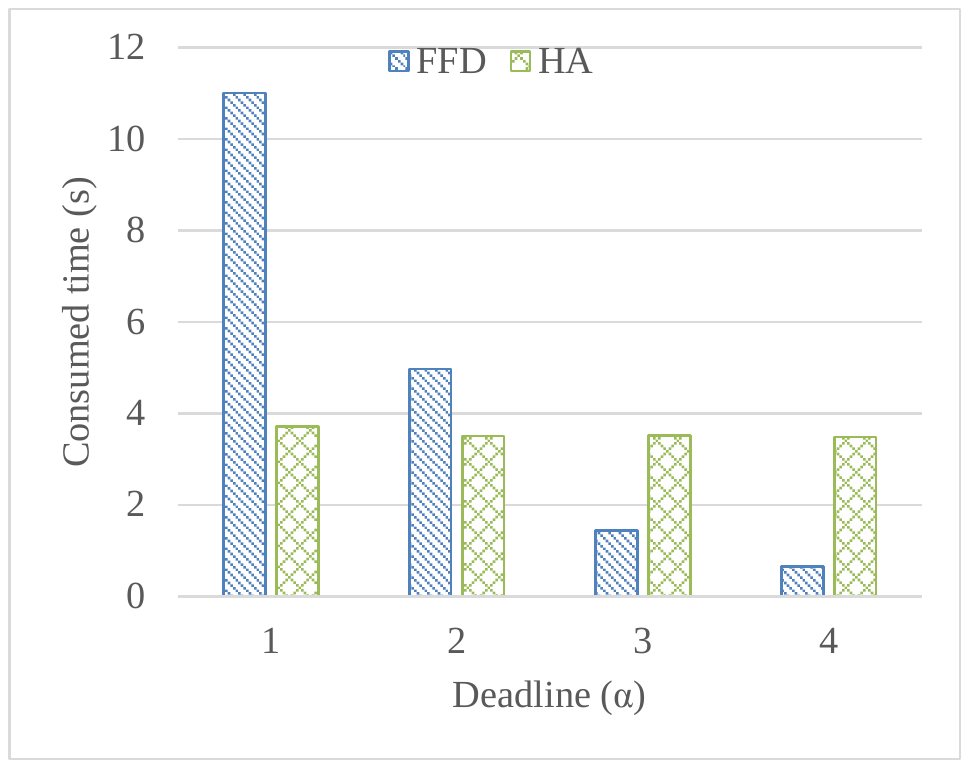}
    }
    \caption{{The time consumed by task assignments.}}\label{time}
\end{figure}

\begin{figure}[!t]
    \centering
    \subfloat[Gaia]{
    \includegraphics[width=0.45\columnwidth]{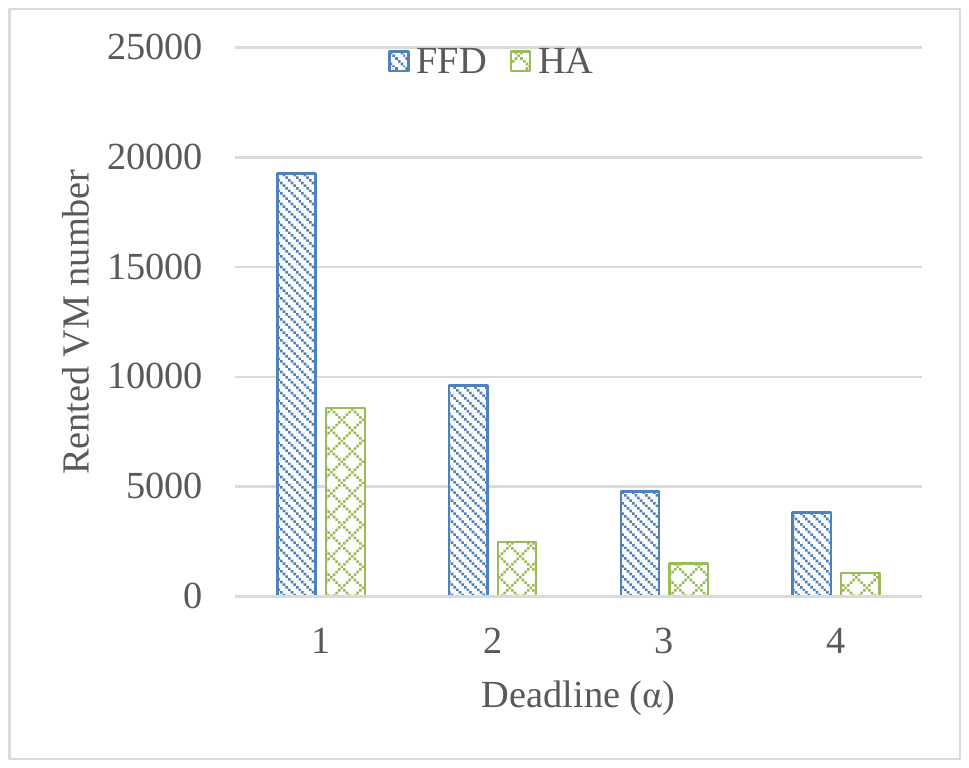}
    }
    \subfloat[NASA]{
    \includegraphics[width=0.45\columnwidth]{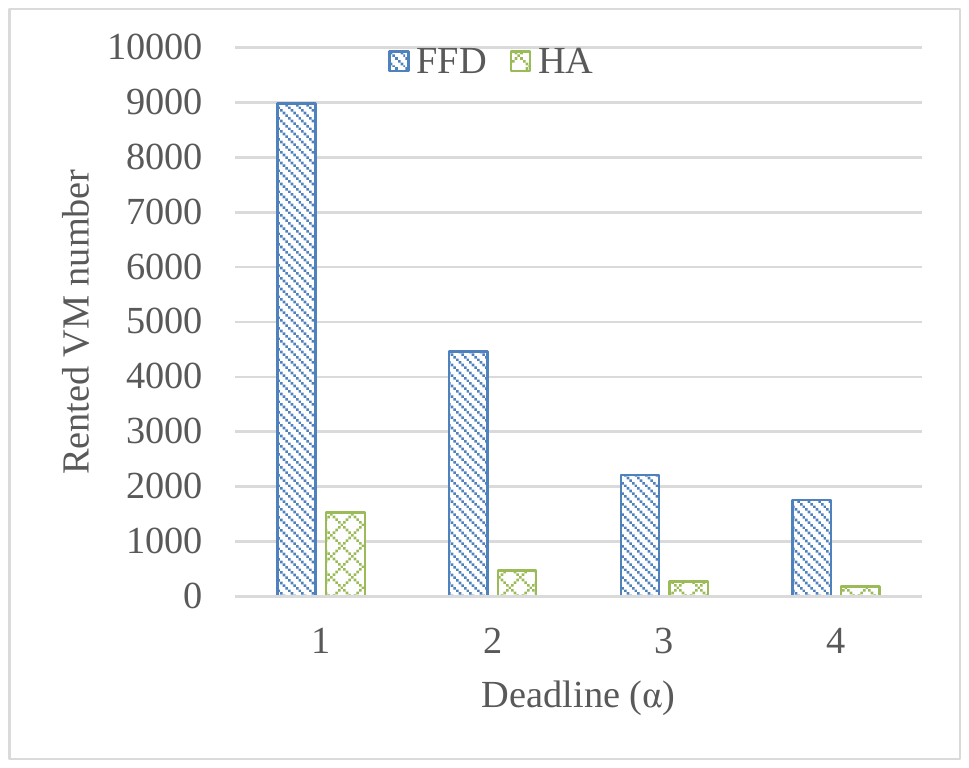}
    }
    \caption{{The numbers of rented VMs for completing tasks within respective deadlines.}}\label{VMnumber}
\end{figure}

Figure~\ref{time} shows the time consumed by HA and FFD running on a Intel(R) Xeon(R) CPU E5410 @ 2.33GHz core. As shown in this figure, HA consumes less time than FFD when the deadline is early ($\alpha=1/2$), while it consumes more time than FFD when it is late ($\alpha=3/4$). The reasons are as follows. The time complexities of HA and FFD are $O(T\cdot(P+T+N_{RV}))$ and $O(T\cdot(P+N_{RV}))$, respectively. When the deadline is earlier, the number of rented VMs is larger. When $\alpha=1$ or 2, the number of rented VMs used by FFD is larger than 4000, as shown in Fig.~\ref{VMnumber}, which is above 288\% more than that by HA and is larger than the number of tasks for both Gaia and NASA traces, and thus the time consumed by FFD is more than HA. When $\alpha\geq 3$, rented VM numbers are smaller than task numbers, therefore, FFD consumes less time than HA.

\begin{figure}[!t]
    \centering
    \includegraphics[width=0.8\columnwidth]{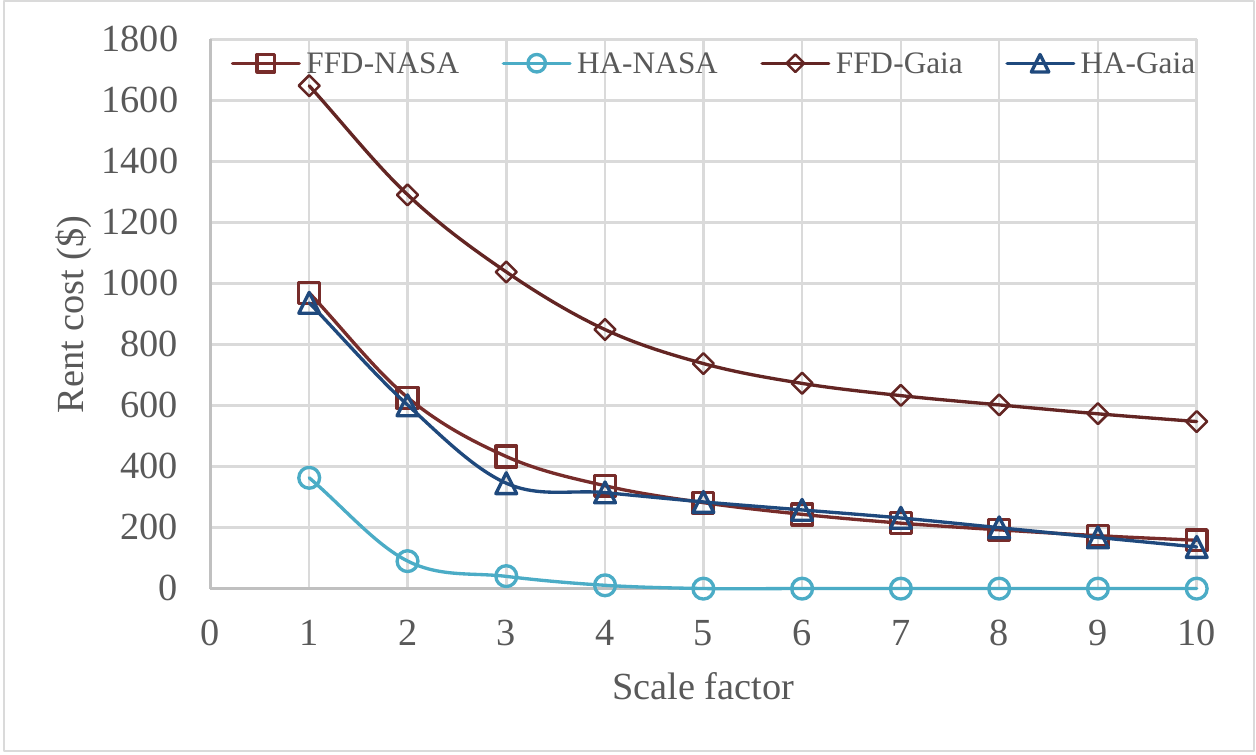}
    \caption{{The rent costs decreasing with increasing of PM numbers.}}\label{scale1}
\end{figure}

\begin{figure}[!t]
    \centering
    \includegraphics[width=0.8\columnwidth]{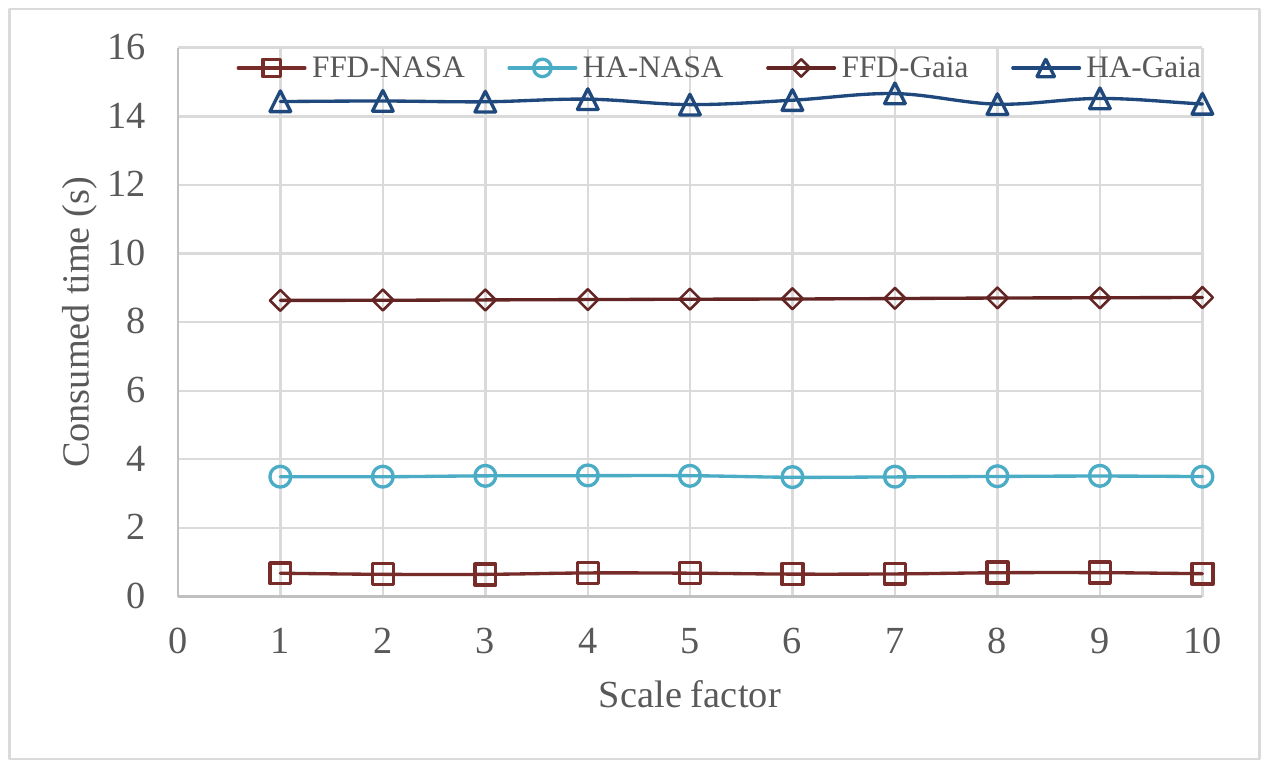}
    \caption{{The times consumed by task assignments, stable when the local PM number is increasing.}}\label{scale2}
\end{figure}

Next, we examine the scalabilities of HA and FFD on the number of local PMs. We scale the PM resources by a factor ranging from 1 to 10. For example, when the scale factor is 1, 15 PMs described in Table~\ref{PM} are used for running tasks, while there would be 150 PMs when the scale factor is 10.
 Figures~\ref{scale1} and \ref{scale2} show the changes of costs and time consumed by HA and FFD with the scale factor, respectively. We present the results of the case of $\alpha=4$ here. Other cases have similar results. As shown in Fig.~\ref{scale1}, the cost is decreased with increasing of PM numbers because the resources should be leased from the public cloud are reduced in amount as the amount of local resources increases. From Fig.~\ref{scale1}, we can see that HA costs less 24.1\%-75.1\% and about 20\% than FFD for Gaia and NASA traces, respectively, for any scale of local cloud, i.e., HA always consumes less cost than FFD. As shown in Fig.~\ref{scale2}, the times consumed by FFD and HA both are stable as the increase of PM number. The reason is that the increase of PM number results in the decreasing of rented VM number, leading to the total number of the PMs and rented VMs ($P+N_{RV}$) having almost no change as the PM number increases.
 
\begin{figure}[!t]
    \centering
    \includegraphics[width=0.8\columnwidth]{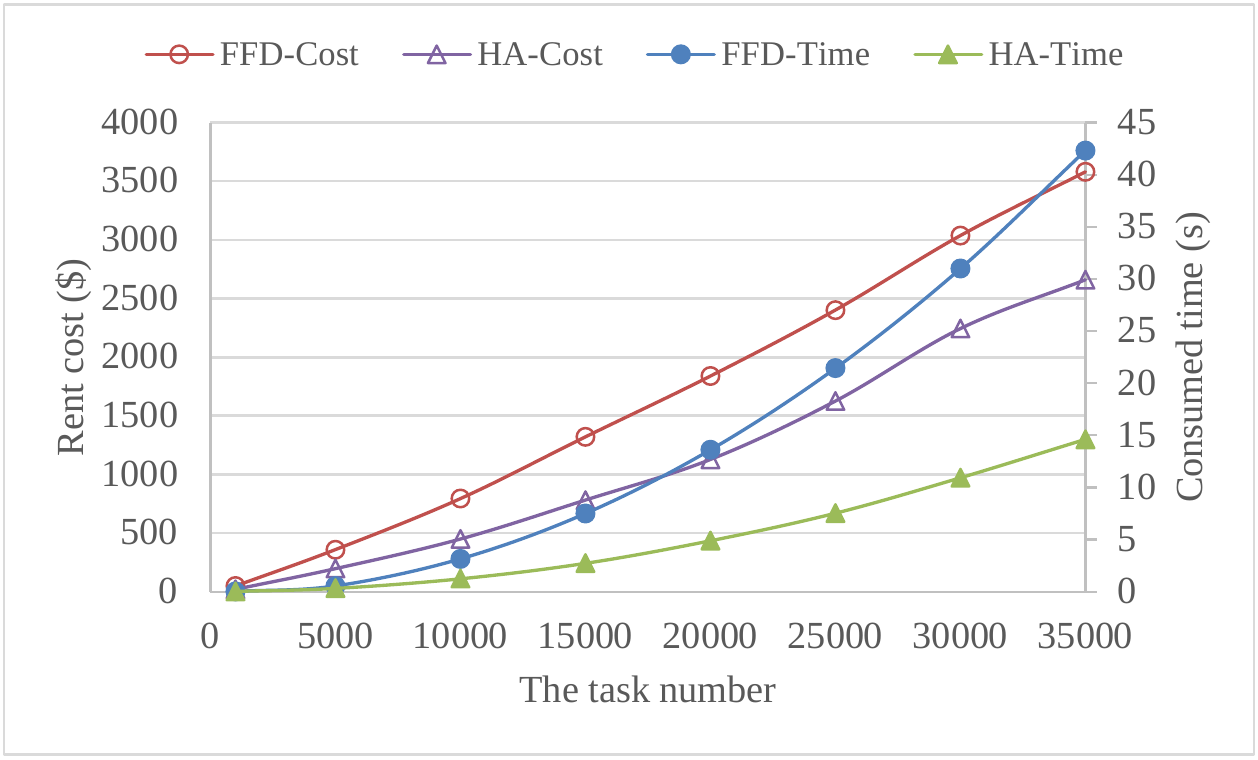}
    \caption{{The variations of rented costs and consumed times by task assignments with increasing the task number.}}\label{scaleTask}
\end{figure}
 
Now, we examine the scalabilities of HA and FFD on the number of tasks by changing the task scale from 1000 to 35000 in number using Gaia trace. Figure~\ref{scaleTask} presents the result of the case of $\alpha=1$. Other cases have similar results. As shown in the figure, the consumed times of HA and FFD both are quadratically increasing with the task number, which is consistent with their respective time complexities. Figure~\ref{scaleTask} shows that the costs have nearly linear increases with the increase of task number as more resources required for processing more tasks and that HA saves 25.7\%-58.1\% costs compared with FFD.

These results above indicate that both HA and FFD have good scalability and that HA is better than FFD in minimizing rent cost.

\section{Related Work} \label{RW}

Outsourcing jobs to a public cloud is a cost-effective way to address the problem of satisfying the peak resource demand when the local cloud/cluster has insufficient resources. There are various researches on scheduling scientific applications on hybrid clouds.

To minimize the cost for leased resources from public clouds, W. Z. Jiang and Z. Q. sheng \cite{CSC} modelled the mapping of tasks and VMs as a bipartite graph. The two independent vertex sets of the bipartite graph are task and VM collections, respectively. The weight of an edge is the VM cost of a discrete task, i.e. the product of the running time of the task and the cost of the VM per unit time. Then the problem minimizing the cost is to find a subset of the edge set, where the weighted sum of all the edges in the subset is the minimum. The authors used the Hopcroft-Karp algorithm \cite{HKA} to solve the minimum bipartite match problem. This work does not consider whether a task could be finish within its deadline.

Some existing work studied on minimizing cost with deadline constraints in hybrid clouds. Van den Bossche et al. \cite{BOT1,BOT2,BOT3} proposed a set of algorithms to cost-efficiently schedule the deadline-constrained BoT applications on both public cloud providers and private infrastructure while taking into account data constraints, data locality and inaccuracies in task runtime estimates.
The Hybrid Cloud Optimized Cost (HCOC) scheduling algorithm\cite{HCOC1,HCOC2} tried to optimize the monetary execution costs resulting from the public nodes only while maintaining the execution time fitting deadline. HCOC first made an initial schedule using the Path Clustering Heuristic (PCH) algorithm \cite{PCH} to schedule the tasks to the private cloud and then rescheduled some tasks to the public cloud if the deadline is missed. The algorithm can achieve cost optimization for workflow scheduling. 
Genez et al. \cite{UCC13} presented an integer linear program model. The numbers of each type of VM instances in both private and public clouds are obtained by solving this model to minimize cost without missing its deadline for a workflow. In this work, the authors considered the VM instances with same type being homogeneous, which is not apply to a private cloud with heterogeneous PMs.
For completing the job on time and with minimum cost, Chu and Simmhan \cite{reusable} first modelled a time-vary spot VM price as a Markov chain and then established a reusable table with three-tuple elements consisted of job compute requirement, deadline constraint of the job and a series of actions with minimal cost based on the price model. The subsequent action was decided by performing a simple table look-up. In this work, they considered that utilizing off local machines does not incur expense.

These above work studied on either resource provisioning or task scheduling. Moreover, most approaches for dynamic provisioning operate in a per-job level, and thus they are inefficient because they fail in consider that other tasks could utilize idle cycles of cloud resources. To address these problems, Aneka \cite{Aneka1,Aneka2,Aneka3} coordinated dynamic provisioning and scheduling that is able to cost-effectively complete applications within their respective deadlines by considering the whole organization workload at individual tasks level when making decisions and an accounting mechanism to determine the share of the cost of utilization of public cloud resources to be assigned to each user. While Aneka separately scheduled tasks and provisioned resources, i.e., Aneka first decided how many resources, each of which is either a VM in public cloud or a PM in local cloud/grids, used for running tasks and then provisioned the resources from the resource pool, considering that all of the resources are homogeneous.

Besides minimizing cost, a few work focused on minimizing the makespan of scientific applications by cloud bursting. FermiCloud \cite{FermiCloud} despatched a VM on the PM that has the highest utilization but still have enough resource for the VM in private cloud. Only when all the resources in private cloud are consumed, VM are deployed on a public cloud. A new VM would be launched in a public cloud only when adding the VM can reduce the average job running time. Kailasam et al. \cite{CBS1,CBS2} proposed four cloud bursting schedulers whose main ideas are outsourcing a job to a public cloud when the estimated time between now and beginning execution of the job is greater than the estimated time consumed by migrating the job to the public cloud.

These aforementioned work studied on the task and/or resource management with one objective minimizing financial cost for private cloud providers or minimizing the makespan of applications. There are a few work focusing on a balance between two objectives.
Taheri et al. \cite{bio} proposed a bi-objective optimization model minimizing both the execution time of a batch of jobs and the transfer time required to deliver their required data for hybrid clouds, and used a PSO-based approach to find the relevant pareto frontier. They do not take the finance expenditure into account.
V. A. Leena et al. \cite{GABIO} proposed an algorithm for the simultaneous optimization of execution time and cost, in hybrid cloud, by determining whether tasks have to be scheduled to either the private cloud or the public cloud, employing a genetic algorithm. This work does not consider the mapping between VMs and PMs in the private cloud. The VM instances of the same type are homogeneous, which is not true in a heterogeneous data center.
Wang et al. \cite{AsQ} proposed a dual-objective multi-dimension multichoice knapsack problem to model the task scheduling with the two objectives of minimizing the cost and minimizing the makespan in hybrid clouds. As the high complexity of solving the problem, the adaptive scheduling algorithm (AsQ) was proposed. AsQ used MAX-MIN strategy \cite{MAXMIN} to schedule task in private cloud and outsourcing the smallest task to the public cloud when the private cloud has insufficient resources. AsQ allocated the public resource slot with minimal cost to a task, fitting the deadline constraint of the task, considering that using extra public resource slots does not incur extra expense. HoseinyFarahabady et al.\cite{Pareto1,Pareto2,Pareto3} studied on the balance between the makespan and the cost for BoT applications in hybrid clouds. They first established a BNP model with the objective of minimizing the sum of the weighted costs, i.e., the product of cost per unit time of a task and the running time raised to the power of a predefined factor of the task, and then relaxed the model by removing the binary constraints. By Lagrange multiplier method, the relaxed model was solved to get the workload assigned to each resource (PM in the private cloud or VM in public clouds). At last, they used FFD algorithm \cite{FFD} to assigned tasks to resources so that the total workload of a resource are close to the value obtained from the last step.

Different from all of the above work, we study on cost-efficiently mapping the tasks to the resources for deadline-constrained BoT applications on a hybrid cloud with heterogeneous local resources.

\section{Conclusion}\label{CFW}

In this paper, we study on the management of BoT jobs with deadline constraints on hybrid clouds. To minimize the cost of VMs leased from public cloud to complete jobs within their respective deadlines, respectively, we model the task schedule and resource provision into a BNP problem. As the BNP problem is NP-hard, we propose a heuristic algorithm to solve the problem in polynomial time. The main idea of the heuristic algorithm is assigning the task whose finish time is closest to and no later than its deadline if running on the current PM or rented VM. if there is no such assignment, the algorithm add a PM with maximum available capacity or a VM with best cost-performance ratio when there is no available PM.
We conduct extensive experiments using two real work traces to investigate the effectiveness and efficiency of the proposed algorithm. The experiments results show that our heuristic algorithm saves 16.2\%-76\% cost for finishing jobs within their respective deadlines and improves 47.3\%-182.8\% resource utilizations, compared with first fit decreasing (FFD) algorithm, with good scalability.

\section*{ACKNOWLEDGMENT}
The authors are grateful to the anonymous reviewers' comments. This work was supported in part by the project of NSFC under grants 61202060, 912183001, 61173112 and 61221062, the National High-Tech Research and Development Program (863) of China under grant 2013AA01A212.

\small
\bibliographystyle{IEEEtran}
\bibliography{IEEEabrv,BoT-SpringSim}

\end{document}